\documentclass[aps,prl,superscriptaddress,twocolumn,nofootinbib]{revtex4-2}

\usepackage{amsfonts}
\usepackage{amsmath}
\usepackage[normalem]{ulem}
\usepackage{amssymb}
\usepackage{mathtools}
\usepackage{graphicx}
\usepackage{enumerate}
\usepackage{wrapfig}
\usepackage{color}
\usepackage{yhmath}
\usepackage{mathrsfs}
\usepackage{bbm}
\usepackage{commath}
\usepackage[hidelinks]{hyperref}
\hypersetup{
     colorlinks   = true,
     citecolor    = blue,
     linkcolor    = blue
}
\usepackage{subfigure}
\usepackage{mathptmx}
\usepackage{times}
\usepackage{ulem}

\DeclareMathAlphabet{\pazocal}{OMS}{zplm}{m}{n}

\DeclareMathOperator{\tr}{tr}

\DeclareMathSymbol{\Alpha}{\mathalpha}{operators}{"41}
\DeclareMathSymbol{\Beta}{\mathalpha}{operators}{"42}
\DeclareMathSymbol{\Epsilon}{\mathalpha}{operators}{"45}
\DeclareMathSymbol{\Zeta}{\mathalpha}{operators}{"5A}
\DeclareMathSymbol{\Eta}{\mathalpha}{operators}{"48}
\DeclareMathSymbol{\Iota}{\mathalpha}{operators}{"49}
\DeclareMathSymbol{\Kappa}{\mathalpha}{operators}{"4B}
\DeclareMathSymbol{\Mu}{\mathalpha}{operators}{"4D}
\DeclareMathSymbol{\Nu}{\mathalpha}{operators}{"4E}
\DeclareMathSymbol{\Omicron}{\mathalpha}{operators}{"4F}
\DeclareMathSymbol{\Rho}{\mathalpha}{operators}{"50}
\DeclareMathSymbol{\Tau}{\mathalpha}{operators}{"54}
\DeclareMathSymbol{\Chi}{\mathalpha}{operators}{"58}
\DeclareMathSymbol{\omicron}{\mathord}{letters}{"6F}

\def\XXint#1#2#3{{\setbox0=\hbox{$#1{#2#3}{\int}$}
\vcenter{\hbox{$#2#3$}}\kern-.5\wd0}}

\makeatletter
\DeclareFontFamily{OMX}{MnSymbolE}{}
\DeclareSymbolFont{MnLargeSymbols}{OMX}{MnSymbolE}{m}{n}
\SetSymbolFont{MnLargeSymbols}{bold}{OMX}{MnSymbolE}{b}{n}
\DeclareFontShape{OMX}{MnSymbolE}{m}{n}{
    <-6>  MnSymbolE5
   <6-7>  MnSymbolE6
   <7-8>  MnSymbolE7
   <8-9>  MnSymbolE8
   <9-10> MnSymbolE9
  <10-12> MnSymbolE10
  <12->   MnSymbolE12
}{}
\DeclareFontShape{OMX}{MnSymbolE}{b}{n}{
    <-6>  MnSymbolE-Bold5
   <6-7>  MnSymbolE-Bold6
   <7-8>  MnSymbolE-Bold7
   <8-9>  MnSymbolE-Bold8
   <9-10> MnSymbolE-Bold9
  <10-12> MnSymbolE-Bold10
  <12->   MnSymbolE-Bold12
}{}

\newcommand{\ignore}[1]{}
\newcommand{\nobibentry}[1]{{\let\nocite\ignore\bibentry{#1}}}

\newcommand{\bibfnamefont}[1]{#1}
\newcommand{\bibnamefont}[1]{#1}
\newcommand{\ket}[1]{\left\vert#1\right\rangle}
\newcommand{\bra}[1]{\left\langle#1\right\vert}

\newcommand{\bea}{\begin{eqnarray}}
\newcommand{\eea}{\end{eqnarray}}

\renewcommand*{\thefootnote}{\fnsymbol{footnote}}

\newcommand*{\blue}[1]{\textcolor{black}{#1}}

\begin{document}

\title{Page-curve-like entanglement dynamics in open quantum systems}

\author{Jonas Glatthard}
\affiliation{Department of Physics and Astronomy, University of Exeter, Exeter EX4 4QL, United Kingdom}
\email{J.Glatthard@exeter.ac.uk}

\begin{abstract}
The entanglement entropy of a black hole, and that of its Hawking radiation, are expected to follow the so-called Page curve: After an increase in line with Hawking's calculation, it is expected to decrease back to zero once the black hole has fully evaporated, as demanded by unitarity. Recently, a simple system-plus-bath model has been proposed which shows a similar behaviour. Here, we make a general argument as to why such a Page-curve-like entanglement dynamics should be expected to hold generally for system-plus-bath models at small coupling and low temperatures, when the system is initialised in a pure state far from equilibrium. The interaction with the bath will then generate entanglement entropy, but it eventually has to decrease to the value prescribed by the corresponding mean-force Gibbs state. Under those conditions, it is close to the system ground state. We illustrate this on two paradigmatic open-quantum-system models, the exactly solvable harmonic quantum Brownian motion and the spin-boson model, which we study numerically. In the first example we find that \blue{the intermediate entropy of an initially localised impurity is higher for more localised initial states}. In the second example, for an impurity initialised in the excited state, the Page time--when the entropy reaches its maximum--occurs when the excitation has half decayed.
\end{abstract}

\maketitle

\noindent \textit{Introduction.} Entanglement has become an object of intense study in diverse fields from quantum information to quantum many-body and black-hole physics. A central result is that the von Neumann entropy of subregions in the ground state of many-body Hamiltonians with local interactions grows like the surface area of the subregion and not its volume \cite{eisert2010}. Computational matrix-product approaches \cite{cirac2021} use such entanglement properties for efficient simulation of quantum systems, from quantum many-body systems \cite{schollwock2011} to non-Markovian open quantum systems \cite{strathearn2018}. Another important result is that the entanglement entropy of ergodic systems out of equilibrium generically grows linearly until it saturates at a value given by the volume law for excited states \cite{calabrese2005,ho2017,kim2013}. From this perspective, the behaviour of the entropy of black holes is unusual.

By considering quantum fields in curved spacetimes, Hawking found that black holes can evaporate by emitting thermal radiation, thus associating a temperature proportional to the inverse mass to black holes \cite{hawking1975}. This, together with an associated entropy proportional to the area of the event horizon \cite{bekenstein1972}, makes certain laws of black hole mechanics look like the laws of thermodynamics \cite{wald2001}. By considering the situation in which matter in a pure state undergoes gravitational collapse to form a black hole and then evaporates according to Hawkings semiclassical calculation, one finds that a pure state evolves into a mixed state in a closed system, which stands in stark contrast to the unitarity of time evolution in quantum mechanics. This is the so-called black hole information paradox \cite{almheiri2021}. If unitarity were to hold for quantum gravity, instead of a linear increase of entropy according to Hawkings calculation, the curve describing the entropy should bend down and decrease back to zero once the black hole has fully evaporated, so that the final state remains pure \cite{page1993}. The resulting curve is called the Page curve. Recently \cite{almheiri2019, penington2020}, tremendous progress has been made calculating this behaviour semiclassically, using so-called quantum extremal surfaces \cite{ryu2006prl,ryu2006jhep}, first developed in the context of the AdS/CFT correspondence \cite{maldacena1999}.

Recently \cite{kehrein2023}, a relatively simple solvable system-plus-bath model \cite{bp, weiss1999} has been proposed, which shows Page-curve-like entanglement dynamics. \blue{The entanglement entropy is examined as function of time and not as function of subsystem size, as often studied in quantum many-body systems \cite{page1993,calabrese2009,lydzba2020, bianchi2021, bianchi2022}.} Similar observations have been made for the system-bath mutual information \cite{ptaszynski2022} and the entanglement negativity \cite{ptaszynski2024} of open quantum systems. Furthermore, in \cite{kehrein2023} a more general argument was made, that such an entanglement dynamics is generally expected when the resulting open dynamics forces the system into a low dimensional subspace of its Hilbert space. In the example given, the fermionic system empties out.

Here we make an argument that the condition suggested in \cite{kehrein2023} can be generically realised for open quantum systems weakly coupled to an environment at low temperature. When initialised in a pure far-from-equilibrium state, the ensuing non-equilibrium dynamics will be accompanied by a high entanglement entropy production. Under our conditions, however, one can expect, that the impurity will eventually settle down to a state close to its ground state, which carries a low entanglement entropy. The entanglement entropy as a function of time will therefore qualitatively look like the Page curve. We will corroborate and illustrate this on paradigmatic open-system models; namely, the exactly solvable harmonic quantum Brownian motion \cite{caldeira1983path, caldeira1983tunnelling, hanggi2005, lampo2019, cresser2023}
and the spin-boson model \cite{lambert2019,boudjada2014,yang2014,purkayastha2020,thoss2001,anders2007}, which we study numerically.

The main ingredient of our argument is the so-called mean-force Gibbs state \cite{cresser2021, timofeev2022, trushechkin2022}, which amounts to the reduced state of the global system-bath canonical equilibrium state. It is conjectured \cite{trushechkin2022} that open quantum systems in large enouth thermal baths generically approach this steady state. The physical picture behind this is the following. As the environment is much larger than the impurity, the state of the impurity is just a perturbation to the global equilibrium and the dynamics essentially consists of the return to equilibrium of the global system, in line with the eigenstate thermalisation hypothesis \cite{deutsch1991,srednicki1994,alessio2016,gogolin2016}. Under our conditions, that is, weak coupling and low temperature, the mean-force Gibbs state is close to the local ground state and thus carries low von Neumann entropy.\\

\noindent \textit{Entanglement dynamics.} We study the entanglement dynamics for an impurity, the system, embedded in an environment, the bath. The total Hamiltonian for such a problem takes the form commonly studied in the open quantum systems literature
\begin{equation}\label{eq:h_tot}
	\pmb H = \pmb H_S + \pmb H_B + \pmb H_I,
\end{equation}
where the subscripts $S,B$ and $I$ stand for system, bath and interaction. Here, and in what follows, operators will be denoted by boldface symbols. Considering the situation where the probe is initialised in a pure state $\pmb \rho_0$ and the environment is in its ground state (which we assume to be non-degenerate), the global product state is also pure. Despite the global state remaining pure under the dynamics generated by the Hamiltonian in Eq.~\eqref{eq:h_tot}, the subsystems (impurity and environment) get entangled as they interact. We quantify this by the entanglement entropy, i.e., the von Neumann entropy $S$ of the subsystems given by
\begin{equation} \label{eq:SvN}
	S = - \tr \left( \pmb \rho \log \pmb \rho \right).
\end{equation}
The entanglement entropy of the system and that of its complement (the bath) are the same. Generically, for large systems avoiding recurrence, is it expected that the entanglement entropy rises with time and reaches a plateau asymptotically \cite{calabrese2005,ho2017,kim2013}.

On the other hand, when the bath is much larger than the system, it is generically expected that the open system reaches a steady state given by the so-called mean-force Gibbs state \cite{cresser2021, timofeev2022, trushechkin2022}
\begin{equation} \label{eq:mfgs}
	\pmb{\tau}_{MF} = \frac{\tr_B e^{-\beta\,\pmb{H}}}{\tr e^{-\beta\,\pmb{H}}},
\end{equation}
where $\beta$ is the inverse temperature of the bath. This is the expectation as long as no symmetries lead to conserved quantities constraining the system dynamics. Here, and in what follows, we work in natural units, i.e. $\hbar = k_B = 1$. This means that the system reaches a state which looks like the reduction of the system-bath equilibrium at the temperature of the bath.  Our environment being in its ground state, this means that we expect a steady state for the impurity taking the form $\pmb{\tau}_{MF}(T=0) = \tr_B \ket{\Omega} \bra{\Omega}$, where $\ket{\Omega}$ is the global ground state.

At very weak coupling between system and bath, as commonly assumed in thermodynamics and, more generally, whenever the coupling is given by a surface term, while the system and bath energies are given by large volume terms, the mean force ground state can be approximated well by the system ground state $\ket{0}$, i.e. $\pmb \rho_S(t \rightarrow \infty) = \tr_B \ket{\Omega} \bra{\Omega} \simeq \ket{0} \bra{0}$. The entanglement entropy, therefore, must be close to the one of the local ground state, which we assume to be non-degenerate. This means the entanglement entropy must be close to zero. Importantly, this has to hold for arbitrary initial states, in particular states far from equilibrium whose dissipative dynamics result in large entropy production.

Putting our pieces together we have to following picture: Impurity and environment start at a pure state with zero entanglement entropy. The entanglement entropy can increase substantially as they interact during the non-equilibrium dynamics. At long times the entanglement entropy needs to reach the value of the mean-force state, which is close to zero. In between it has to decay, therefore qualitatively following the Page curve.

The above is very similar to evaporation of AdS black holes. This is usually facilitated by coupling them, via a surface term, to an auxiliary CFT in the vacuum state \cite{almheiri2019}. Those are essentially the conditions we have described here.

In the following we study the entropy dynamics quantitatively on concrete models. Despite working with small coupling, we do this without using weak coupling master equations \cite{davies1974, gks1976, lindblad1976} prevalent in the open quantum systems literature, as they are not applicable at such low temperatures \cite{bp}. In particular we do not assume Markovianity. Instead we work with exact solutions.\\

\noindent \textit{Harmonic quantum Brownian motion.} The first example is the exactly solvable harmonic quantum Brownian motion, i.e. a harmonic oscillator linearly coupled to a continuum of harmonic oscillators. This model can, for example, describe a Bose polaron in a condensate \cite{lampo2017}. The system Hamilton reads
\begin{equation}\label{eq:sys-ho}
	\pmb H_S = \frac{1}{2} \omega_R^2\,\pmb x^2 + \frac{1}{2}\,\pmb p^2,
\end{equation}
where we have set the mass to $1$.
The bath Hamiltonian is given by 
\begin{equation}\label{eq:h_bath}
    \pmb H_B = \sum\nolimits_\mu \omega_\mu^2 m_\mu \pmb x_\mu^2/2  + \pmb p_\mu^2/(2 m_\mu),
\end{equation}
the most common model for the environment in the open quantum systems literature. Such an environment could represent the electromagnetic field or phonons in a crystal. Finally the coupling between system and bath is given by
\begin{equation}
	\pmb H_I = \pmb x\otimes\sum\nolimits_\mu g_\mu \pmb x_\mu.
\end{equation}
The effect of the bath on the system can be encoded via the spectral density $J(\omega) = \pi\,\sum\nolimits_\mu g_\mu^2/(2 m_\mu \omega_\mu)\,\delta(\omega - \omega_\mu)$. We choose the common Ohmic spectral density with Lorentz--Drude cutoff given by 
\begin{equation}\label{eq:sd}
	J(\omega) = \frac{\gamma\,\omega}{1 + (\omega/\Lambda)^2}.    
\end{equation}
This spectral density is linear for frequencies much smaller than the cutoff $\Lambda$ and decays for larger frequencies. Other choices of spectral densities would lead to qualitatively similar results. We choose the frequency $\omega_R$ in Eq.~\eqref{eq:sys-ho} to cancel out the distortion due to the bath with regard to the oscillator frequency $\omega_0$ by setting
\begin{equation}
	\omega_R^2 = \omega_0^2 +  \Delta\omega^2    
\end{equation}
where the counter-term frequency is given by $\Delta\omega^2 = \frac{2}{\pi}\int_{0}^\infty d\omega\,J(\omega)/\omega$. Such a counter-term naturally arises in many systems of interest \cite{caldeira1983tunnelling}. For our spectral density it evaluates to $\Delta\omega^2 = \gamma\Lambda$.

\begin{figure}[t]
\centering
\includegraphics[width=0.4\textwidth]{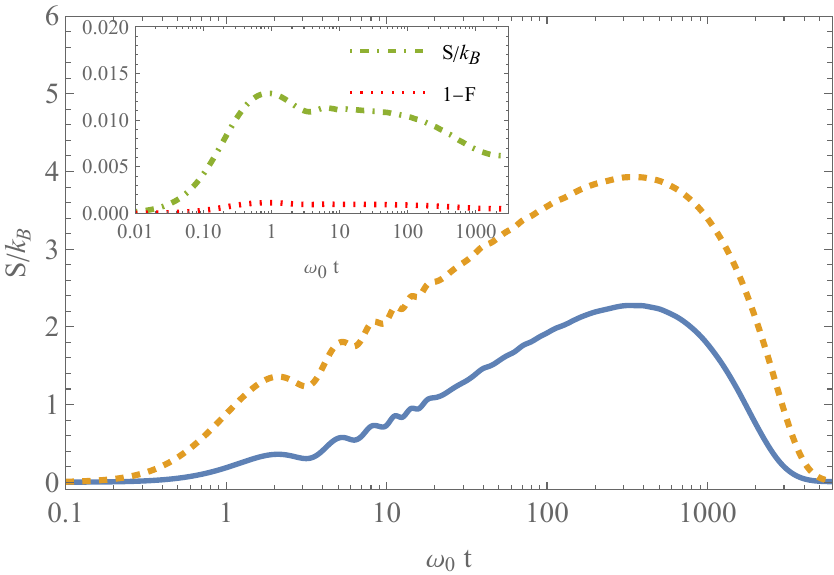}
\caption{\blue{The entanglement entropy $S$ of the oscillator impurity as a function of time starts at zero, as the impurity is initialised in a pure state, for a localised wave packet with width $\delta = 1/100$ (solid blue) and $\delta = 1/1000$ (dashed orange). The interaction entangles the impurity with the environment, which is initially at zero temperature. Their entanglement entropy peaks at an intermediate time. The more localised the initial state, the higher the intermediate peak of the entanglement entropy. Regardless of initial state, the mean-force Gibbs state is approached at long times, which, for weak coupling, is close to the local ground state. As in the Page curve, the entanglement entropy thus has to decrease; here, it saturates close to zero. In the inset the initial state is the local ground state, the maximum of the entanglement entropy (dot-dashed green) still reaches an intermediate maximum, but it is just slightly higher than the final value, as the dynamics is not very far from equilibrium, seen by the fidelity $F$ with the local ground state staying close to one (dotted red). The parameters are $\omega_0=1, \gamma=0.001$ and $\Lambda=10$.}}
\label{fig1}
\end{figure}

Initialising the bath in a thermal state, which is Gaussian, we exploit the fact that the Hamiltonian is quadratic in positions and momenta, thus generating a \textit{Gaussianity}-preserving dynamics \cite{ferraro2005}. That is, for Gaussian initial conditions with vanishing first moments, the covariances $ \sigma_{xx} = \langle \pmb{x}^2 \rangle $, $ \sigma_{xp} = \frac12\langle \pmb{x}\pmb{p} + \pmb{p}\pmb{x} \rangle$ and $ \sigma_{pp} = \langle \pmb{p}^2 \rangle $ fully characterise the state of the impurity. We collect the covariances in the matrix
\begin{equation}
    \mathsf{\Sigma} = \begin{pmatrix} \sigma_{xx} & \sigma_{xp} \\ \sigma_{xp} & \sigma_{pp} \end{pmatrix}.
\end{equation}
In the Supplemental Material \cite{SM} we give details on the exact dynamics of the model.

\begin{figure}[t]
\centering
\includegraphics[width=0.4\textwidth]{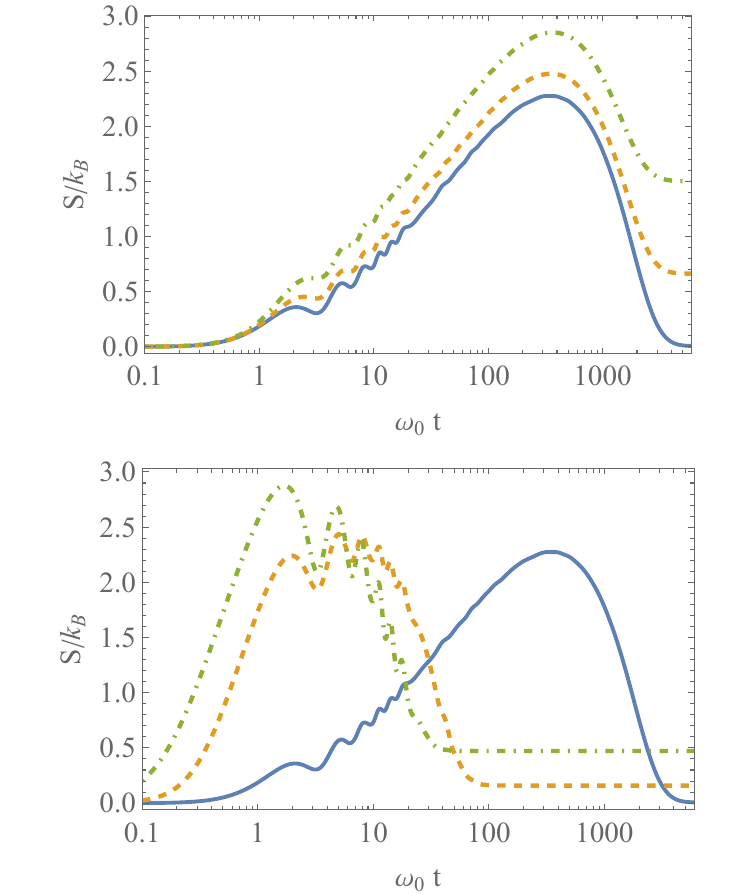}
\caption{\blue{\textbf{(top)} The entropy of the oscillator impurity, initialised in a localised wave packet with with $\delta = 1/100$, reaches an intermediate maximum not just at absolute zero, i.e. $T=0$ (solid blue), but also when the bath is initialised at a finite, but low temperature, here $T=0.5$ (dashed orange) and  $T=1$ (dot-dashed green). \textbf{(bottom)} Dependence of the entropy dynamics on the strength of the coupling, for $\gamma=0.001$ (solid blue), $\gamma=0.05$ (dashed orange) and $\gamma=0.1$ (dot-dashed green). The overall qualitative features of the curve do not change. The value of the entropy at the intermediate peak is of the same order for the different coupling strengths. In contrast, the weaker the coupling, the closer to zero is the asymptotic value. The other parameters are as in Fig.~\ref{fig1}.}}
\label{fig2}
\end{figure}

For Gaussian states we can express the von Neumann entropy, Eq.~\eqref{eq:SvN}, as \cite{demarie2012}
\begin{equation} \label{eq:SG}
    S = \left(\lambda + \frac{1}{2}\right) \log_2 \left( \lambda + \frac{1}{2}\right)-\left(\lambda - \frac{1}{2}\right) \log_2 \left( \lambda - \frac{1}{2}\right),
\end{equation}
where $\lambda$ is the symplectic eigenvalue of the state. This is calculated as the the absolute value of the eigenvalues $\{i \lambda , - i \lambda \}$ of the matrix product $\Sigma \Omega$, with the symplectic matrix 
\begin{equation}
\Omega = \begin{pmatrix} 0 & 1 \\ -1 & 0 \end{pmatrix}.
\end{equation}

\begin{figure}[t]
\centering
\includegraphics[width=0.4\textwidth]{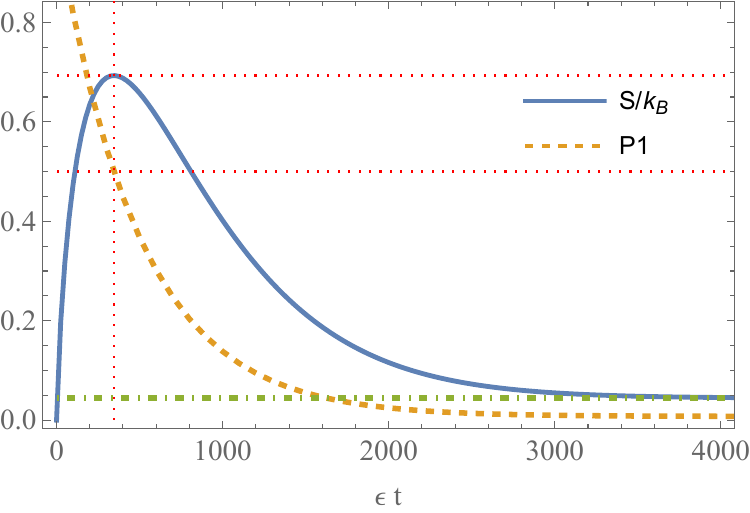}
\caption{The entropy $S$ (solid blue) of a two-level impurity as a function of time. The impurity is initialised in the excited state, the bath is at a temperature close to zero. No \blue{off-diagonals of the reduced state in the impurity energy basis} are generated in the dynamics and the excited state population $P1$ (dashed orange) monotonously decays. Therefore the Page time corresponds to the time at which the ground state and the excited state are equally populated (dotted red lines). The entropy \blue{of this state is} the maximum possible value for a two-level system, \blue{i.e. $ln(2)$}. In the long time limit the entropy converges to the entropy of the mean-force Gibbs state (dot-dashed green), which is close to zero, thus behaving qualitatively like the Page curve. \blue{If initialised in the impurity ground state, the entanglement entropy approaches is final value without an intermediate peak (not shown).} The parameters are $\epsilon=1, \gamma=0.001, \Lambda=10$ and $T=0.2$.}
\label{fig3}
\end{figure}

\begin{figure}[t]
\centering
\includegraphics[width=0.4\textwidth]{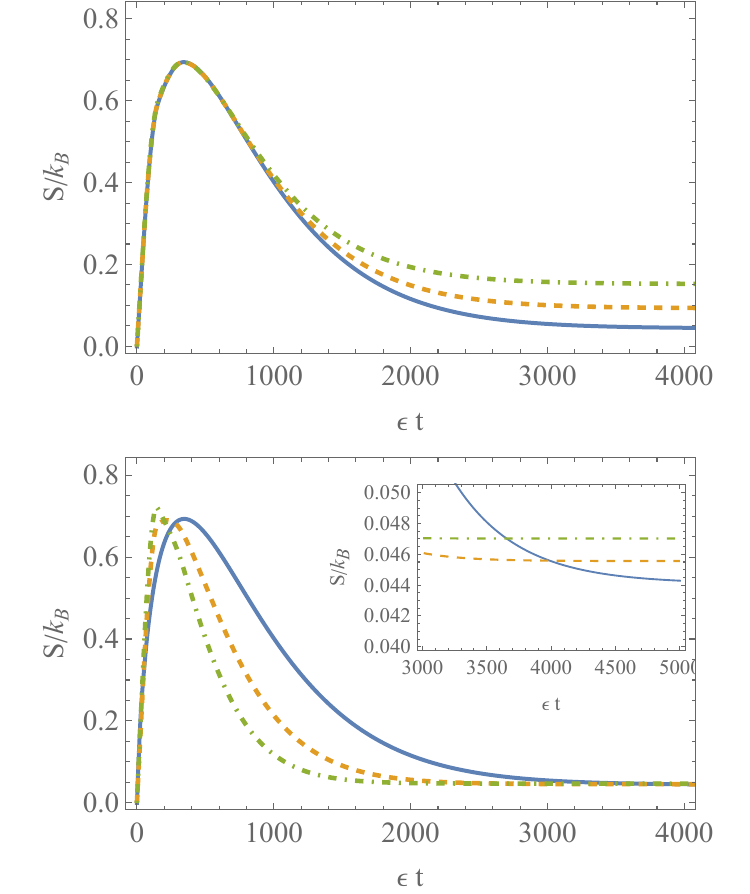}
\caption{\textbf{(top)} The entropy of the two-level impurity, again initialised in the excited state, as a function of time at different environment temperatures, i.e. $T=0.2$ (solid blue), $T=0.25$ (dashed orange) and $T=0.3$ (dot-dashed green) behaves qualitatively the same \blue{and reaches the maximum entropy as long as the temperature is low enough}, only the asymptotic values are different. The lower the temperature the closer to zero the asymptotic value. \textbf{(bottom)} Dependence of the entropy dynamics on the overall strength of the coupling to the environment, here by varying the parameter $\gamma$, for $\gamma=0.001$ (solid blue), $\gamma=0.0015$ (dashed orange) and $\gamma=0.002$ (dot-dashed green). For smaller coupling the excited state decays slower, leading to a later peak in entropy and the asymptotic value gets closer to zero, as shown in the inset. The other parameters are as in Fig.~\ref{fig3}.}
\label{fig4}
\end{figure}

\blue{We consider impurity initial states which are Gaussian wave packets highly localised at the origin. We will quantify the position variance as $\sigma_{xx}=\delta$ and the momentum variance $\sigma_{pp}=0.5^2/\delta$ and take $\sigma_{xp}=0$. Those are pure states and saturate the Heisenberg inequality. The environment is initialised in its ground state.} The global state is pure and therefore the von Neumann entropy is just the entanglement entropy. In Fig.~\ref{fig1} we show the entanglement entropy, calculated by Eq.~\eqref{eq:SG}, as a function of time. We observe that the entropy, starting from zero as the impurity is initialised in a pure sate, reaches a maximum at intermediate times due to the interaction and then decays back to the entropy of the mean-force Gibbs state, a value close to zero. \blue{The intermediate peak of the entanglement is higher, the more localised the initial state is. Here, even if initialised in the local ground state and staying close to it, the intermediate peak is slightly higher than the final value. The closeness is calculated by the Uhlmann fidelity, which can easily be calculated for Gaussian states \cite{scutaru1998}.}

In Fig.~\ref{fig2} we study the robustness of the entropy dynamics of the impurity against changes of parameters, by varying the environment temperature and \blue{coupling strength}. In the case of non-zero temperature, the global state is not pure and the von Neumann entropy is therefor not just the entanglement entropy. We observe that for low temperatures the qualitative behaviour of the entropy as a function of time stays the same, but changes when the temperature becomes larger. At high enough temperature, the entropy of the impurity does not decay to a value close to zero, but continues to grow and saturate at a higher value. Again for an environment initially in its ground state, changing the strength of the interaction leaves the qualitative picture unchanged, \blue{as long as it stays small}. The lower the coupling, the lower the final value of the entropy. Reducing the strength of the interaction, the final value of the entanglement entropy can get arbitrarily close to zero.\\

\noindent \textit{Spin-boson model.} We now turn our attention to a case study for which no exact solution is available—the spin–boson model, i.e. a two-level system coupled to the continuum of harmonic oscillators. This model can, for example, describe an exciton interacting with phonons \cite{nazir2016}. The system Hamiltonian in this case reads
\begin{equation}\label{eq:sys-qb}
	\pmb H_S = \frac{1}{2} \epsilon \pmb \sigma_z,
\end{equation}
and the coupling between system and bath is given by
\begin{equation}
	\pmb H_I = \pmb \sigma_x \otimes\sum\nolimits_\mu g_\mu \pmb x_\mu,
\end{equation}
where the $\sigma_\alpha$ are the Pauli matrices.
Like in the previous example the bath is given by Eq.~\eqref{eq:h_bath} and the spectral density by Eq.~\eqref{eq:sd}. As the impurity couples to the environment via $\sigma_x$, for which $\sigma_x^2 = \pmb 1$, no counter-term is needed as there is no distortion of the system's potential \cite{cresser2021,timofeev2022,cerisola2022,correa2023}.

We turn to numerical methods to solve the dynamics; namely, we use the numerically exact hierarchical equations of motion (HEOM) approach \cite{tanimura1989, tanimura2020}. We use the QuTiP-BoFiN implementation \cite{lambert2023},  which is integrated in the QuTiP platform \cite{johansson2012, johansson2013}. To numerically calculate the dynamics using HEOM we have to truncate two expansions. One is the Matsubara expansion of the bath correlation function in an exponential series, we call the number of retained terms $N_k$. The other is the level of the hierarchy for the auxiliary density operators, which we call $N_C$. To speed up convergence, the Tanimura terminator has been used \cite{lambert2023}. Our calculations have converged for $N_k = 30$ and $N_C = 2$, which we confirmed by comparing against calculations with higher $N_k$ and $N_C$.

The two-level impurity is initialised in the pure excited state with density matrix $\pmb \rho_0 = \ket{1} \bra{1}$ and the environment in a thermal state at a temperature close to zero. The latter is done for computational reasons, as for lower temperatures the computations are numerically more demanding. The von Neumann entropy is therefore not just the entanglement entropy, as the global state is not pure.

In Fig.~\ref{fig3} we show the von Neuman entropy, Eq.~\eqref{eq:SvN}, as a function of time. We observe that the entropy, starting from zero as the impurity is initialised in a pure sate, reaches a maximum at intermediate times and then decays back to a value close to zero. No off-diagonals in the impurity energy basis are generated during the time evolution. The population of the excited state, also shown in the Figure, monotonously decays to its final value. The time at which the impurity has an equal population in the excited state and ground state is when the entropy reaches its maximum and starts to decay, i.e., the Page time. The entropy at this time is the maximum possible entropy for a two-level system, \blue{i.e. $ln(2)$}.

In Fig.~\ref{fig4} we study the parameter dependence of the entropy dynamics of the impurity by varying the environment temperature and the coupling strength. We observe that for lower temperatures the final value of the entropy is closer to zero. As the global state is also closer to being a pure state, the von Neumann entropy is closer to the entanglement entropy. Similarly, the lower the coupling between impurity and environment, the lower the final value of the entropy. At the Page time, the entropy reaches the maximum allowed value for a two-level system for all coupling strengths.\\

\noindent \textit{Conclusion.} In this letter we studied the entropy dynamics of quantum impurities weakly coupled to an environment close to the absolute zero. We gave a general argument why the entropy as a function of time should qualitatively look like the Page curve when the impurity is initialised in a pure state far from equilibrium. For two paradigmatic open-system models, we quantitatively illustrated this effect without making any approximations. We found that for a \blue{localised} oscillator impurity, the entanglement entropy show a Page-curve-like behaviour, \blue{with the peak value of the entropy depending on how localised the impurity initially was}. For a two-level impurity initialised in the excited state we found that the entropy at the Page time reaches the maximally possible value independently of coupling strength, when the excitation has half decayed. Further, we found the effect robust to changes in coupling strength as long as the overall coupling stays small and even for small non-zero temperatures of the environment.

The systems considered here are simple enough to study in detail. In particular, a discretised approximation of the environment could be followed for the whole evolution. This gives a possibility to study of how the environment becomes more pure after the Page time and retains a memory of the system initial state. The study of such simple toy models might therefore open up new paths to study analogues to the purification of the Hawking radiation after the Page time.

Furthermore, the conditions under which our findings hold are very general and should be suitable for experimental realisation. In particular, the studied models are commonly realised in cold atoms and solid state physics.\\

\noindent \textit{Acknowledgements.} The author gratefully acknowledges useful discussions with Luis A. Correa and Stefan Kehrein. This work is supported by a scholarship from CEMPS at the University of Exeter.\\

\renewcommand{\thefootnote}{\fnsymbol{footnote}}

\bibliographystyle{apsrev4-2}
\bibliography{references}

\newpage

\widetext
\section{Supplemental material}

The Heisenberg equations of motion for the oscillator impurity can be written as the quantum Langevin equation \cite{weiss1999} 
\begin{align}
	\ddot{\pmb x}(t) + \omega_R^2\,\pmb x(t) - \int_0^t d\tau\,\chi(t-\tau)\, \pmb x(\tau) = \pmb F(t), \label{eq:QLE}
\end{align}
where $\chi(t)$ is the `dissipation kernel'
\begin{equation}
	\chi(t) = \frac{2}{\pi} \int_0^\infty J(\omega) \sin{(\omega t)}\,d \omega,
\end{equation}
and $\pmb{F}(t)$ is the so-called quantum stochastic force, which takes the form
\begin{align}
	\pmb F(t) = - \sum_\mu g_\mu \left( \pmb x_\mu(0) \cos{(\omega_\mu t)} + \frac{\pmb p_\mu(0)}{m_\mu \omega_\mu} \sin{(\omega_\mu t)} \right).
\end{align}

Introducing the notation $ \pmb{\mathsf{Z}} = (\pmb x, \pmb p)^\mathsf{T} $ allows to write the exact solution of Eq.~\eqref{eq:QLE} as \cite{correa2023}
\begin{subequations}\label{eq:G-t}
	\begin{equation}
		\pmb{\mathsf{Z}}(t) = \mathsf{G}(t)\,\pmb{\mathsf{Z}}(0) + \int_0^t \mathsf{G}(t-t')\,\pmb{\mathsf{F}}(t')\,dt',
	\end{equation}
with entries of $\mathsf{G}(t)$
	\begin{align}
		[\mathsf{G}(t)]_{11} &= [\mathsf{G}(t)]_{22} =\pazocal{L}_t^{-1}\left[\frac{s}{s^2 + \omega_R^2 - \widehat{\chi}(s)}\right],\\
		[\mathsf{G}(t)]_{12} &=\pazocal{L}_t^{-1}\left[\frac{1}{s^2 + \omega_R^2 - \widehat{\chi}(s)}\right]=: g(t),\\
		[\mathsf{G}(t)]_{21} &=\pazocal{L}_t^{-1}\left[\frac{\widehat{\chi}(s)-\omega_R^2}{s^2 + \omega_R^2 - \widehat{\chi}(s)}\right],
	\end{align}
\end{subequations}
and $\pmb{\mathsf{F}}(t) = (0,\pmb F(t))^T$. Here, $\widehat{f}(s):=\int_0^\infty dt\,e^{-s\,t}\,f(t)$ stands for the Laplace transform and $\pazocal{L}^{-1}_t[\,g\,]$ for the inverse transform. For our spectral density, we have $\widehat{\chi}(s) =  \frac{\gamma \Lambda^2}{s + \Lambda}.$

With this, we can calculate the time evolution of the covariance matrix as
\begin{equation}
	\mathsf{\Sigma}(t) = \textrm{Re}\,\langle \pmb{\mathsf{Z}}(t)\,\pmb{\mathsf{Z}}(t)^\mathsf{T} \rangle - \langle \pmb{\mathsf{Z}}(t)\rangle \langle \pmb{\mathsf{Z}}(t)^\mathsf{T}\rangle. \label{covmatrix}
\end{equation}
In the long-time limit we get
\begin{subequations}\label{eq:covariances_coefficients}
	\begin{align}
		&\sigma_{xx}(t \rightarrow \infty) =\frac{1}{\pi}\int_0^\infty \vert \widehat{g}(i \omega )\vert^2\,J(\omega)\,\coth{\frac{\omega}{2T}}\,d\omega, \label{eq:xx_ss}\\
		&\sigma_{pp}(t \rightarrow \infty) =\frac{1}{\pi}\int_0^\infty \omega^2\,\vert\widehat{g}(i \omega )\vert^2\,J(\omega)\,\coth{\frac{\omega}{2T}}\,d\omega,  \label{eq:pp_ss}\\
		&\sigma_{xp}(t \rightarrow \infty) =0 \label{eq:xp_ss}.
	\end{align}
\end{subequations}

\end{document}